\documentclass[english]{IEEEtran}
\usepackage[T1]{fontenc}
\usepackage{color}
\usepackage{float}
\usepackage{mathrsfs}
\usepackage{amsmath}
\usepackage{graphicx}

\makeatletter

\providecommand{\tabularnewline}{\\}
\floatstyle{ruled}
\newfloat{algorithm}{tbp}{loa}
\providecommand{\algorithmname}{Algorithm}
\floatname{algorithm}{\protect\algorithmname}

\usepackage[caption=false,font=footnotesize]{subfig}
\usepackage{algorithm,algpseudocode}
\usepackage{amsmath}

\@ifundefined{showcaptionsetup}{}{%
 \PassOptionsToPackage{caption=false}{subfig}}
\usepackage{subfig}
\AtBeginDocument{

}

\makeatother

\usepackage{babel}
\begin{document}

\title{Two-dimensional Burst Error Correcting Codes using Finite Field Fourier Transforms}

\author{Shounak Roy and~Shayan Garani Srinivasa\\
 Department of Electronic Systems Engineering\\
Indian Institute of Science, Bengaluru: 560012, India\\
Email: \{shounak,shayan.gs\}@dese.iisc.ernet.in}
\maketitle
\begin{abstract}We construct two-dimensional codes for correcting burst errors using
 the finite field Fourier transform. The encoding procedure is performed
in the transformed domain using the conjugacy property of the finite
field Fourier transform. The decoding procedure is also done in the
transformed domain. Our code is capable of correcting multiple non-overlapping
occurrence of different error patterns from a finite set of predefined
error patterns. The code construction is useful for encoding data
in two dimensions for application in data storage and bar codes.
\end{abstract}


\section{Introduction}

Data storage technologies in magnetic and flash memories are shifting towards a two-dimensional (2D)
paradigm since it offers better signal-to-noise (SNR) ratio and format efficiency \cite{key-1},
 inturn leading to higher storage densities. Examples of such storage technologies
include 2D magnetic recording \cite{key-2}, bit-patterned media \cite{key-3}, flash storage \cite{key-4},
optical holographic recording \cite{key-5} etc. The native
format of data applicable to these technologies is inherently two-dimensional. This is useful for
combating 2D intersymbol-interference, noise and other impairments. Storage channels have noise bursts along with an intersperse of random errors \cite{key-6}. The data to be encoded and written onto the storage medium must be capable of overcoming a mixture of burst and random errors. Thus, designing efficient 2D error correcting codes along with modulation codes is important for these storage channels along with sophisticated signal processing for timing recovery, equalization and detection \cite{key-7} as a precursor step before decoding. In this paper, we develop a theory for encoding 2D data resilient to burst errors. 

While there are many well known approaches for correcting
1D burst errors, the design of codes for correcting 2D errors is non
trivial for the following reasons: (a) The shape of a 2D burst can be arbitrary, unlike a 1D burst string
of errors. (b) The errors can be correlated in 2D and non-separable, requiring a 2D
frame work for efficient encoding and decoding. In the following sub-sections, we will survey 2D coding techniques relevant within the context of our work and highlight our contribution.

\subsection{Prior work and context}

The fundamental theory of 2D cyclic codes was first formulated by
Imai \cite{key-8} with the construction of generator and parity check
tensors for 2D codewords. The construction was analogous to the construction
of 1D linear block codes. This code could correct only a single burst error
of an arbitrary error pattern and was not designed for correcting
single burst of more than one error pattern. In a recent paper by
Yoon and Moon \cite{key-9}, they provided the construction of a code
with disjoint syndrome sets among specific error patterns, improving
the construction in \cite{key-8}. This helped in correction of a
single occurrence of an error pattern from a finite set of different
error patterns. However, they did not address the correction of multiple
burst errors.\textit{\textcolor{red}{{} }}Others \cite{key-10}-\cite{key-11}
have constructed codes for correcting a small cluster of errors by
converting the 2D binary codewords into symbols of non-binary codewords
by grouping the bits along rows and columns. Although they can correct
an error pattern of any shape, their method could not correct multiple
disjoint cluster of errors. Blahut \cite{key-12} introduced higher
dimensional finite field transforms for realizing algebraic codes
over curves. Madhusudhana and Siddiqui \cite{key-13} extended Blahut's
decoding algorithm for 2D BCH codes. Analogous to the 1D BCH code,
the 2D extension is required to have a block of contiguous positions
set to zero in the frequency domain, where, the syndromes are calculated
sequentially. They show correction of random and burst errors from
a deconvolution viewpoint. However, their work does not incorporate
the idea of disjoint syndrome sets which otherwise can lead to miscorrections.
Shiozaki \cite{key-14} presented a new class of 2D codes over $GF(q)$
using 2D Fourier transform techniques of dimension $(q-1)^{2}$. In
that work, an exact relationship between the parity check symbols
and the minimum distance was also provided. However, the code construction
does not accommodate disjoint syndrome sets. Also, the details of
the decoding procedure are not explicitly provided. In this paper, we present a construction
of 2D codewords in the frequency domain for correcting multiple
burst errors. Our work is based on the 2D finite field transform inspired from the
work in \cite{key-12}.

\subsection{Our approach and contribution}

We present a construction of 2D linear codes using the finite field Fourier
transform for correcting multiple non-overlapping occurrences of
different error patterns from a finite set of predefined error patterns.
Our encoding scheme uses the conjugacy property of the 2D finite field Fourier transform. Unlike prior approaches \cite{key-8}, the encoding process is simple as it does not need any parity check or generator
tensors. Further, our code construction is tailored
for correcting a set of predefined error patterns as well as an arbitrary linear combination from the set of predefined error patterns.
Most 1D decoding approaches follow a two step process of identifying error locations followed by error correction. In the 2D case, we first need to identify which error pattern
has occurred. With this information, we try to solve the erroneous
locations. In case we do not identify the error pattern, we declare
an occurrence of uncorrectable error. In our work, we also consider the correction
of multiple disjoint bursts of the same pattern assuming the number
of bits to be corrected falls within the minimum distance bound, a feature not
considered in \cite{key-9}-\cite{key-11}. The set of common roots
among all the codewords in the code forms the set of common zeros
\cite{key-8}. To facilitate the decoding procedure, we formulate
the common zero set accordingly. A subset of the common zero set is
defined as the set of indicator zeros. This set of zeros are used
to identify the error pattern that might have occurred. The remaining
set of zeros are used to form the syndrome equations in the transformed
domain for solving the erroneous locations in contrast to the
approach in \cite{key-15,key-16}. The entire common zero set is also
chosen in a way to obtain disjoint syndrome sets among a set of predefined
error patterns/error events. Yoon and Moon \cite{key-9} gave a criterion to obtain
disjoint syndrome sets among specific error patterns. We interpret
the same criterion from a transform domain perspective. The construction in \cite{key-13} required a block of contiguous zeros in the transform domain. However, in our work, we relax this constraint
and choose the zeros for \textit{a-priori} identification of error
patterns, and obtaining disjoint syndrome sets among predefined error
patterns. 

The paper is organized as follows. In Section II, we highlight the
code description and motivate the 2D finite field Fourier transform
(FFFT). In Section III, we develop an encoding scheme for mapping
2D binary arrays into 2D codewords using FFFT. In Section IV, we discuss
the decoding of 2D codewords in the transform domain that can correct
simultaneous occurrence of disjoint error patterns with illustrative examples. We conclude the
paper in Section V.

\section{Two dimensional Binary Cyclic Codes }

\subsection{Code Description }

A 2D binary code $\mathscr{\mathscr{C}}$ of dimension $n\times m$
is a collection of arrays having $n$ rows and $m$ columns with elements
from $GF(2)$. A 2D array can be described using a
bi-variate polynomial expression as,

\begin{equation}
c(x,y){\displaystyle =\sum_{(i,j)\in\Omega}}c_{i,j}x^{i}y^{j},
\end{equation}

where, $\Omega=\{(i,j)|0\leq i\leq n-1,\mathrm{}0\leq j\leq m-1\}$
and $c_{i,j}\in GF(2)$. 

A 2D code is cyclic if $x^{i}y^{j}c(x,y)\in\mathscr{\mathscr{C}}\text{, }\forall c(x,y)\in\mathscr{\mathscr{C}}$.
A 2D binary code is said to be linear if $\sum_{i}c_{i}(x,y)\in\mathscr{\mathscr{C}}\text{, }\forall c_{i}(x,y)\in\mathscr{\mathscr{C}}$
. Let $\mathscr{C}[\Omega]$ denote the set of
all bi-variate polynomials. If $f(x,y)\in\mathscr{C}[\Omega]$, then
\begin{equation}
\{f(x,y)\}_{\Omega}=f(x,y)+a(x,y)(x^{n}-1)+b(x,y)(y^{m}-1),
\end{equation}

where, $a(x,y)$ and $b(x,y)$ are polynomials over $GF(2)$. Consider
a bi-variate polynomial $f(x,y)$. A point $(x',y')$ is said to be
a root of $f(x,y)$ if $f(x',y')=0.$ For a code of dimension $n\times m$,
let $\gamma$ and $\beta$ be the primitive $n^{th}$ and $m^{th}$
roots of unity respectively i.e., $\gamma^{n}=1$ and $\beta^{m}=1.$
All roots constitute the set $\mathcal{V}$ given by $\mathcal{V}=\{\left(\gamma^{i},\beta^{j}\right)|0\leq i\leq n-1,\mathrm{}0\leq j\leq m-1\}.$
The roots common to all the codewords in the code are referred to as the
set of \textit{common zeros}. For any set of zeros $\varDelta$, common
to some arbitrary polynomials over $GF(2)$, if $\left(\delta,\sigma\right)\in\Delta$
then $\left(\delta^{2^{k}},\sigma^{2^{k}}\right)\in\Delta$ for $k=1,2,...,n-1$
and $n$ is the least positive integer for which $\left(\delta,\sigma\right)=\left(\delta^{2^{n}},\sigma^{2^{n}}\right)$.
These points are known as conjugate points in $\Delta$. From the
set $\mathcal{V}$, we select a subset $\mathcal{V}_{c}$, which will
be our set of common zeros. A 2D code is completely characterized
by this set of common zeros.

\subsection{2D Finite Field Fourier Transform (FFFT) }

The 2D finite field Fourier transform (FFFT) for an array of dimension
$n\times m$ having elements from GF(2) is defined as 
\begin{equation}
C_{\theta,\phi}=\sum_{i=0}^{n-1}\sum_{j=0}^{m-1}\gamma^{i\theta}\beta^{j\phi}c_{i,j}.\label{eq:ffft}
\end{equation}

$C_{\theta,\phi}\in GF(2^{\lambda})$, $2^{\lambda}=\mathrm{{lcm}}(n,m)+1$,
where $\mathrm{lcm(.)}$ denotes the least common multiple. If we
have $n=3$ and $m=5$, the elements of $C_{\theta,\phi}$ are from
$GF(2^{4})$. Thus, a binary codeword gets converted to a non-binary
codeword of the same dimension. This is a direct extension from the
1D FFFT defined in \cite{key-12}

\subsection{2D Finite Field Inverse Fourier transform}

The 2D finite field inverse finite field Fourier transform (IFFFT)
is defined as
\begin{equation}
c_{i,j}=\frac{1}{\mathrm{mod}(n,p)}\frac{1}{\mathrm{mod}(m,p)}\sum_{\theta=0}^{n-1}\sum_{\phi=0}^{m-1}\gamma^{-i\theta}\beta^{-j\phi}C_{\theta,\phi},\label{eq:iffft}
\end{equation}

where, $p$ is the characteristic of the field and $\mathrm{mod}(n,p)$
is defined as the remainder obtained while $n|p.$

\section{Encoding }

We highlight the encoding procedure in \cite{key-8} in this section
for completeness. The encoding of the 2D codewords in \cite{key-8}
begins by selecting the common zero set $\mathcal{V}_{c}$ which
is followed by the creation of the parity check tensor. The generator
tensor is then created from the parity check tensor and used for encoding
a message. Later in the section, we describe the encoding process
of the code using 2D FFFT, and highlight its various advantages over
\cite{key-8}.

\subsection{Parity Check Tensor }

The parity check tensor \cite{key-8} can have two forms. In the first
form, each element of the parity check tensor is a binary vector of
length equal to the cardinality of the common zero set. The other
form has a bi-variate polynomial as its elements. The second form
is required because we need to generate our 2D codewords in polynomial
form.

Two zeros of $\mathcal{V}_{c}$ are said to be equivalent if the
first component of their coordinate are conjugates over $GF(2)$. Thus, to find the parity
position, we have to partition $\mathcal{V}_{c}$ into equivalence classes.
Two roots $\left(\epsilon,\eta\right)\text{ and }(\theta,\mu)$ are
said to be equivalent \cite{key-8} i.e., $\left(\epsilon,\eta\right)\sim(\theta,\mu),$
for some positive integer $k$, if $\epsilon^{2^{k}}=\theta$. Each
component of the parity check tensor, which is a binary vector, is
split in length depending on the size of each of the equivalence classes.
For a particular $(i,j)$ coordinate of the tensor, we express $\gamma^{i}\beta^{j}$
in terms of $GF\left(2^{|\mathcal{V}{}_{c_{k}}|}\right)$ for the
$k^{th}$ segment. The binary vectors created at the parity positions
are all linearly independent. The binary vectors at the message positions
can be expressed as a linear combination of these vectors. Let the
set of all coordinates of the 2D code be defined as $\Omega$, and,
the set of all parity positions be defined by the set $\pi$. Each vector
at position $(i,j)$ is expressed as
\begin{equation}
\mathbf{h_{i,j}}=\sum_{(k,l)\in\pi}h_{k,l}^{(i,j)}\mathbf{h_{k,l}},\label{eq:parity1}
\end{equation}

where, $h_{k,l}^{(i,j)}\in GF(2)$ and $(i,j)\in\Omega$. To get the
algebraic form of the parity check tensor, we express the components
in bi-variate polynomial form as
\begin{equation}
h_{i,j}(x,y)=\sum_{(k,l)\in\pi}h_{k,l}^{(i,j)}x^{k}y^{l}.\label{eq:parity2}
\end{equation}

The operators $T_{x}$ and $T_{y}$ \cite{key-8} operating on $h_{i,j}(x,y)$
is defined as
\begin{eqnarray}
T_{x}[h_{i,j}(x,y)] & = & h_{\mathrm{mod(}(i+1),n),j}(x,y),\label{eq:tx}\\
T_{y}[h_{i,j}(x,y)] & = & h_{i,\mathrm{mod((j+1),}m)}(x,y).\label{eq:ty}
\end{eqnarray}
From the construction of the parity check tensor in the algebraic
form, we see that, $h_{0,0}(x,y)$ component is always 1. Few properties
of $T_{x}$ and $T_{y}$ \cite{key-8} are as follows: 
\begin{enumerate}
\item $T_{x}^{i}T_{y}^{j}=h_{i,j}(x,y).$
\item $f(T_{x},T_{y})={\displaystyle \sum_{(i,j)\epsilon\Omega}}f_{i,j}h_{i,j}(x,y).$
\end{enumerate}
A residue of a polynomial $f(x,y)$ is defined as $f(T_{x,}T_{y}).$
A bi-variate codeword $f(x,y)$ is said to be valid if its residue
is zero, i.e., 
\begin{equation}
f(T_{x},T_{y})=0.\label{eq:valid_in_time}
\end{equation}

\subsection{Generator Tensor }

The relation between $(i,j)$ components of the generator tensor and
the parity check tensor is given by, 
\begin{equation}
g_{i,j}(x,y)=x^{i}y^{j}-h_{i,j}(x,y).\label{eq:generator}
\end{equation}
From equation (\ref{eq:valid_in_time}), it is easy to verify that
each component of the generator tensor is a valid 2D codeword. 

Let the message polynomial be 
\[
m(x,y)={\displaystyle \sum_{(i,j)\in\Omega}}m_{i,j}x^{i}y^{j},
\]
 where $m_{i,j}=0$ for $(i,j)\in\pi$.

The work done in \cite{key-8} does not provide an explicit mathematical
expression for encoding. We prove a result on the encoding of a message vector
$m(x,y)$ into a valid codeword $c(x,y)$ in the following Lemma.

\subsubsection*{Lemma 1}

The encoding equation of the message vector $m(x,y)$ into a codeword
$c(x,y)$ follows as, 
\begin{equation}
c(x,y)=\sum_{(i,j)\in\Omega}m_{i,j}g_{i,j}(x,y).\label{eq:time_enc}
\end{equation}

\begin{IEEEproof}
We know that a codeword $c(x,y)$ is valid if $c(T_{x},T_{y})=0$.
So, we can write the codeword $c(x,y)$ as
\begin{eqnarray}
c(x,y) & = & m(x,y)-m(T_{x,}T_{y}).\label{eq:msgres}
\end{eqnarray}

Simplifying further by using properties of $T_{x}$ and $T_{y}$,
and using (\ref{eq:generator}) in (\ref{eq:msgres}), we have
\begin{eqnarray*}
c(x,y) & = & \sum_{(i,j)\in\Omega}m_{i,j}x^{i}y^{j}-\sum_{(i,j)\in\pi}m_{i,j}T_{x}^{i}T_{y}^{j},\\
 & = & \sum_{(i,j)\in\Omega}m_{i,j}g_{i,j}(x,y),
\end{eqnarray*}

thereby, proving the result.
\end{IEEEproof}

\subsection{Finite field transform of a code}

The original form and the transformed representation of codewords
are defined in the time and frequency domains respectively. We use
the FFFT technique to transform the codewords in the frequency domain.
The data is encoded directly into the transformed domain. This procedure
is easier because it helps us to save space for not storing the
parity check and generator tensors, which are of considerable size.
This reduces the space complexity at the transmitter side extensively.
The binary codewords are obtained from the transformed domain using
2D IFFFT. In the transformed domain, the conjugacy constraints are
used to express the coordinates of the transformed codeword.

\subsubsection*{Lemma 2}

For a 2D code over $GF(q)$,
\begin{equation}
C_{\theta,\phi}^{q}=C_{\left(\mathrm{mod}\left(q\theta,n\right)\right),\left(\mathrm{mod}\left(q\phi,m\right)\right)}.\label{eq:conjfreq}
\end{equation}

\begin{IEEEproof}
For an element $k\in GF(q)$, we have $k^{q}=k$. Also, we know in
case of finite fields, for elements $\{a_{1},a_{2},\cdots,a_{n}\}\in GF(q)$,
the following relation holds: 
\[
\left(a_{1}+a_{2}+\cdots+a_{n}\right)^{q}=\left(a_{1}^{q}+a_{2}^{q}+\cdots+a_{n}^{q}\right).
\]

Thus, from (\ref{eq:ffft}),

\begin{eqnarray*}
C_{\theta,\phi}^{q} & = & \sum_{i=0}^{N-1}\sum_{j=0}^{M-1}\gamma^{iq\theta}\beta^{jq\phi}c_{i,j}^{q},\\
 & = & \sum_{i=0}^{N-1}\sum_{j=0}^{M-1}\gamma^{i(q\theta)}\beta^{j(q\phi)}c_{i,j},\\
 & = & C_{\left(\mathrm{mod}\left(q\theta,n\right)\right),\left(\mathrm{mod}\left(q\phi,m\right)\right).}
\end{eqnarray*}

The last step follows since
\begin{eqnarray}
\mathrm{\mathrm{mod}}\left(q\theta,n\right) & = & q\theta,\nonumber \\
\mathrm{mod}\left(q\phi,m\right) & = & q\phi,
\end{eqnarray}

thereby, proving the result.
\end{IEEEproof}

\subsection{Validity of a codeword in the frequency domain }

\subsubsection*{Lemma 3}

Let $\left(\gamma^{\theta},\beta^{\phi}\right)$ be a zero of the codespace.
In the frequency domain, $C_{\theta,\phi}=0$.
\begin{IEEEproof}
We have,

\[
c(x,y)=\sum_{(i,j)\in\Omega}c_{i,j}x^{i}y^{j}.
\]

Plugging in $\left(\gamma^{\theta},\beta^{\phi}\right)$ for $(x,y)$,
we have

\begin{eqnarray}
\sum_{(i,j)\in\Omega}c_{i,j}\gamma^{i\theta}\beta^{j\phi} & = & c\left(\gamma^{\theta},\beta^{\phi}\right),\nonumber \\
\mathtt{\mathrm{or,}}\sum_{(i,j)\in\Omega}c_{i,j}\gamma^{i\theta}\beta^{j\phi} & = & 0.
\end{eqnarray}

From (\ref{eq:ffft}), $C_{\theta,\phi}=0$, which proves our result.
\end{IEEEproof}
Thus, the condition for checking the validity of a codeword in the
frequency domain is

\begin{equation}
C_{\theta,\phi}=0,
\end{equation}

$\text{\ensuremath{\forall\left(\theta,\phi\right)}, where }\left(\gamma^{\theta},\beta^{\phi}\right)\in\mathcal{V}_{c}$.

\subsubsection*{Example 1}

Let the common zero set in the frequency domain be

\[
\mathcal{VF}_{c}=\{(0,0),(1,1),(1,4),(2,2),(2,3)\}.
\]

From Lemma 3, the components in $\mathcal{VF}_{c}$ will be zero. We show one
valid codeword in the transform domain that satisfies Lemma 3 and
the conjugacy constraints. As an example,
\[
C=\left(\begin{array}{ccccc}
0 & \alpha^{13} & \alpha^{11} & \alpha^{14} & \alpha^{7}\\
\alpha^{10} & 0 & \alpha^{6} & \alpha^{9} & 0\\
\alpha^{5} & \alpha^{3} & 0 & 0 & \alpha^{12}
\end{array}\right).
\]

Now, using the 2D IFFFT equation from (\ref{eq:iffft}), we have the
time domain codeword as 
\[
c=\left(\begin{array}{ccccc}
1 & 1 & 1 & 0 & 0\\
0 & 1 & 0 & 0 & 0\\
0 & 0 & 0 & 0 & 0
\end{array}\right).
\]

Our encoding scheme clearly avoids the creation of parity check and
generator tensors to encode a message. Given the considerable size
of these matrices it helps to save space for storing them. Lemmas
2 and 3 are sufficient to encode the code in the frequency domain.
However, codewords are always stored in binary format in storage devices.
This is where the 2D IFFFT operation comes into play. After encoding,
the codeword undergoes the inverse transformation and stored as binary
data.

\section{Decoding }

In this section, we will discuss decoding in the frequency domain.
In time domain, the received codeword $r$, transmitted codeword $c$,
and the error vector $e$ are related by the following equation:
\begin{equation}
\{r\}_{i,j}=\{c\}_{i,j}+\{e\}_{i,j},
\end{equation}

where, $0\leq i\leq n-1$ and $0\leq j\leq m-1$. As FFFT is an linear
operation, the same equality holds. Let $R$ be the received codeword,
$C$ be the transmitted codeword, and $E$ be the error vector in frequency
domain. We have,
\begin{equation}
\{R\}_{\theta,\phi}=\{C\}_{\theta,\phi}+\{E\}_{\theta,\phi}.
\end{equation}
 According to Lemma 3,
\begin{equation}
\{R\}_{\theta,\phi}=\{E\}_{\theta,\phi},\label{eq:refff}
\end{equation}

$\text{\ensuremath{\forall\left(\theta,\phi\right)} where,}\left(\gamma^{\theta},\beta^{\phi}\right)\in\mathcal{V}_{c}.$

When we receive a codeword, the coordinates corresponding to the common
zero set must be zero according to equation (\ref{eq:refff}). So, we
need to find those frequency components of the received
codeword that are equal to those components of the error vector in the
frequency domain. This results in solving fewer syndrome equations
that finally gives us the error locations directly in the time domain. 

In 1D codewords, the burst length is a code design parameter, and the
shape of the burst is inconsequential. The situation is much more complicated in 2D. For example, a three
bit burst error can be considered in three shapes as shown below: 

\begin{figure}[H]
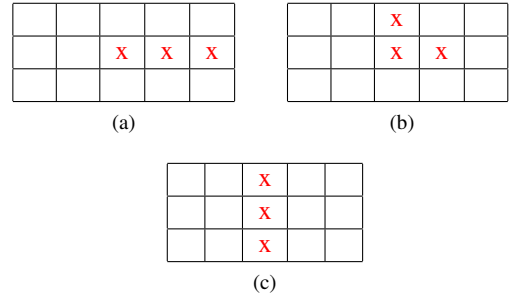

\centering{}\subfloat[]{\centering{}%
\begin{tabular}{|r@{\extracolsep{0pt}.}l|r@{\extracolsep{0pt}.}l|r@{\extracolsep{0pt}.}l|r@{\extracolsep{0pt}.}l|c|}
\hline 
\multicolumn{2}{|c|}{\:\:} & \multicolumn{2}{c|}{\:\:} & \multicolumn{2}{c|}{} & \multicolumn{2}{c|}{} & \tabularnewline
\hline 
\multicolumn{2}{|c|}{\:\:} & \multicolumn{2}{c|}{\:\:} & \multicolumn{2}{c|}{\textcolor{red}{x}} & \multicolumn{2}{c|}{\textcolor{red}{x}} & \textcolor{red}{x}\tabularnewline
\hline 
\multicolumn{2}{|c|}{\:\:} & \multicolumn{2}{c|}{\:\:} & \multicolumn{2}{c|}{} & \multicolumn{2}{c|}{} & \tabularnewline
\hline 
\end{tabular}} \qquad{}\subfloat[]{\centering{}%
\begin{tabular}{|c|c|c|c|c|}
\hline 
 \:\:&\:\:  & \textcolor{red}{x} &\:  &\:\: \tabularnewline
\hline 
 \:\:&\:\:  & \textcolor{red}{x} & \textcolor{red}{x} &\:\: \tabularnewline
\hline 
 &\:\:  &  &\:  &\:\:  \tabularnewline
\hline 
\end{tabular}	} \qquad{}\subfloat[]{\centering{}	%
\begin{tabular}{|c|c|c|c|c|}
\hline 
 \:&  \:& \textcolor{red}{x} &\:  &\: \tabularnewline
\hline 
 \:&  \:& \textcolor{red}{x} &\:  &\: \tabularnewline
\hline 
 \:&  \:& \textcolor{red}{x} &\:  &\: \tabularnewline
\hline 
\end{tabular}			}\protect\caption{(a) A burst error of three bits in horizontal shape. (b) A burst error
of three bits in L-shape. (c) A burst error of three bits in vertical
shape.}
\end{figure}

The code is designed to correct specific error shapes as mentioned
below. The same code construction is extended to correct unknown error
patterns that are a disjoint combination of specific error patterns.
We begin with a few definitions relating error patterns.

\subsubsection*{Definition 1}

A \textit{local 2D error pattern} is defined as an error pattern which
has a polynomial expression $b(x,y)$ over $GF(2)$ such that the coefficient
of the monomial $x^{0}y^{0}$ in $(x,y$) is always 1. For a code
of dimension $n\times m$, the maximum degrees that $x$ and $y$ can take
are $n-1$ and $m-1$ respectively.

\subsubsection*{Definition 2}

A \textit{global 2D error pattern} is defined for a local error pattern
$b(x,y)$ starting from position $(i,j)$. This is given by the following
equation:

\begin{equation}
e(x,y)=x^{i}y^{j}b(x,y).
\end{equation}

As it is clear from Definition 1, an error pattern need not necessarily
mean contiguous erroneous positions. This means that an isolated non-contiguous
erroneous position is also considered as an error pattern.

\subsection{Predefined error patterns}

In this paper, we have considered horizontal and vertical error patterns
as our basic error patterns.

\subsubsection{Error Pattern 1 (Horizontal burst)}

Let us consider a horizontal burst of error of dimension $1\times b_{1}$
starting at position $(i,j)$. The error vector in the time domain
is
\begin{equation}
e_{1}(x,y)=x^{i}y^{j}+x^{i}y^{j+1}+\cdots+x^{i}y^{j+b_{1}}.\label{eq:t12}
\end{equation}
The 2D transformed error vector for (\ref{eq:t12}) is given by
\begin{eqnarray}
E_{\theta,\phi}^{(1)} & = & \sum_{i=0}^{n-1}\sum_{j=0}^{m-1}\gamma^{i\theta}\beta^{j\phi}e_{i,j}=\gamma^{i\theta}\beta^{j\phi}+\gamma^{i\theta}\beta^{(j+1)\phi}+\nonumber \\
 &  & \cdots+\gamma^{i\theta}\beta^{(j+b_{1})\phi}.\label{eq:12}
\end{eqnarray}

The following property holds true.

\subsubsection*{Lemma 4}
A horizontal burst of dimension $1\times(m-1)$ i.e., putting $b_{1}=m$,
cannot be corrected.
\begin{IEEEproof}
From (\ref{eq:12}),
\begin{eqnarray}
E_{\theta,\phi}^{(1)} & = & \gamma^{i\theta}\beta^{j\phi}(1+\beta^{\phi}+\beta^{2\phi}+\cdots+\beta^{\left(b_{1}-1\right)\phi}),\nonumber \\
 & = & \gamma^{i\theta}\beta^{j\phi}(1+\beta+\beta^{2}+\cdots+\beta^{\left(b_{1}-1\right)})^{\phi}.
\end{eqnarray}

As $\beta^{m}=1$, we have,
\begin{equation}
(\beta+1)(1+\beta+\beta^{2}+\beta^{3}+\cdots+\beta^{m-1})=0.\label{eq:z12}
\end{equation}
It is clear from the above equations that the maximum value that $b_{1}$
can take is $m-1$. If $b_{1}=m$, from (\ref{eq:z12}), $E_{\theta,\phi}^{(1)}$
is always zero $\forall(\theta,\phi)\in VF_{c}$ which indicates undetected
error.
\end{IEEEproof}

\subsubsection{Error Pattern 2 (Vertical burst)}

Let us consider a vertical burst of error of dimension $b_{2}\times1$
starting at position $(i,j)$. The error vector in the time domain
is
\begin{equation}
e_{2}(x,y)=x^{i}y^{j}+x^{i+1}y^{j}+\cdots+x^{i+b_{2}}y^{j}.
\end{equation}
The 2D transformed error vector is computed as
\begin{eqnarray}
E_{\theta,\phi}^{(2)} & = & \sum_{i=0}^{n-1}\sum_{j=0}^{m-1}\gamma^{i\theta}\beta^{j\phi}e_{i,j}=\gamma^{i\theta}\beta^{j\phi}+\gamma^{(i+1)\theta}\beta^{j\phi}+\nonumber \\
 &  & \cdots+\gamma^{(i+b_{2})\theta}\beta^{j\phi}.\label{eq:21}
\end{eqnarray}

From Lemma 4 and using $\gamma^{n}=1,$ the maximum value $b_{2}$
can take is $n-1$. 

From here on, we will consider $b_{1}=2$ and $b_{2}=2.$

\subsection{Unknown error patterns}

Simultaneous occurrence of both horizontal and vertical burst errors
could result in a number of unknown error patterns. Considering a horizontal
error pattern of dimension $1\times2$ and a vertical error pattern
of dimension $2\times1$. The following figure shows some possible
unknown error patterns. 

\begin{figure}[H]
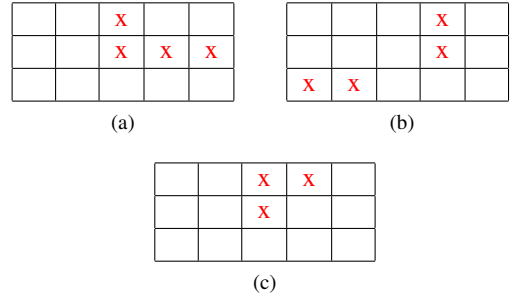

\centering{}\subfloat[]{\centering{}%
\begin{tabular}{|r@{\extracolsep{0pt}.}l|r@{\extracolsep{0pt}.}l|r@{\extracolsep{0pt}.}l|r@{\extracolsep{0pt}.}l|c|}
\hline 
\multicolumn{2}{|c|}{\:\:} & \multicolumn{2}{c|}{\:\:} & \multicolumn{2}{c|}{\textcolor{red}{x}} & \multicolumn{2}{c|}{} & \tabularnewline
\hline 
\multicolumn{2}{|c|}{\:\:} & \multicolumn{2}{c|}{\:\:} & \multicolumn{2}{c|}{\textcolor{red}{x}} & \multicolumn{2}{c|}{\textcolor{red}{x}} & \textcolor{red}{x}\tabularnewline
\hline 
\multicolumn{2}{|c|}{\:\:} & \multicolumn{2}{c|}{\:\:} & \multicolumn{2}{c|}{} & \multicolumn{2}{c|}{} & \tabularnewline
\hline 
\end{tabular}} \qquad{}\subfloat[]{\centering{}%
\begin{tabular}{|c|c|c|c|c|}
\hline 
 &  &\:\:  & \textcolor{red}{x} &\:\: \tabularnewline
\hline 
 &  & \:\: & \textcolor{red}{x} &\:\: \tabularnewline
\hline 
\textcolor{red}{x} & \textcolor{red}{x} &\:\:  &  &\:\: \tabularnewline
\hline 
\end{tabular}	} \qquad{}\subfloat[]{\centering{}	%
\begin{tabular}{|c|c|c|c|c|}
\hline 
 \:\:& \:\: & \textcolor{red}{x} & \textcolor{red}{x} &\:\: \tabularnewline
\hline 
 \:\:& \:\: & \textcolor{red}{x} &  &\:\: \tabularnewline
\hline 
 \:\:& \:\:&  &  &\:\: \tabularnewline
\hline 
\end{tabular}			}\protect\caption{(a) Combination of a $1\times2$ and $2\times1$ error pattern forming
a joint L-shaped burst. (b) Combination of a $1\times2$ and $2\times1$
error pattern forming two disjoint burst error. (c) Combination of
a $1\times2$ and $2\times1$ error pattern overlapping at one position
to form a smaller L-shaped burst.}
\end{figure}

We restrict our code to correct errors with error pattern 1 or error
pattern 2 or a combination of both. Let us assume these error patterns
start at position $\left(k_{1},l_{1}\right)$ and $\left(k_{2},l_{2}\right)$.
Consider the following error polynomial.

\begin{eqnarray}
e(x,y) & = & c_{1}e_{1}(x,y)+c_{2}e_{2}(x,y),\nonumber \\
 & = & c_{1}x^{k_{1}}y^{l_{1}}(1+y)+c_{2}x^{k_{2}}y^{l_{2}}(1+x),\label{eq:totalerroeq}
\end{eqnarray}

where $c_{1},\text{ }c_{2}\in GF(2).$ $c_{1}=1$ and $c_{2}=0$ implies
occurrence of error pattern 1 only. $c_{1}=0$ and $c_{2}=1$ implies
occurrence of error pattern 2 only. $c_{1}=1$ and $c_{2}=1$ implies
occurrence of both error patterns.

\subsection{Selection of common zero set}

The common zero set specifies the code completely. The correction
ability of the code is decided by selecting the common zero set. The
common zero set is selected considering the following two salient
points:
\begin{itemize}
\item Disjoint syndrome sets among predefined error patterns.
\item Inclusion of indicator roots that help us to check the presence of
a particular type of error.
\end{itemize}
We elaborate these steps now.

\subsubsection{The disjoint burst criteria}

According to (\ref{eq:refff}), the syndrome components obtained at
the common zero positions are equal to those corresponding components
of the error vector in the frequency domain. We equate these set of
equations and solve for $i$ and $j$. A criteria for obtaining disjoint
syndrome set is mentioned in \cite{key-9}. If $\mathcal{E}_{1}$
and $\mathcal{E}_{2}$ are the set zeros of two error patterns $b_{1}(x,y)$
and $b_{2}(x,y)$ respectively, and if $\mathcal{V}_{c}$ is the set
of common zeros, then the following result must follow to get disjoint
syndrome set between the aforesaid error patterns i.e., 
\begin{equation}
\mathcal{E}_{1}\cap\mathcal{V}_{c}\neq\mathcal{E}_{2}\cap\mathcal{V}_{c}.
\end{equation}
This can be understood in the frequency domain in a different way.
Consider only horizontal and vertical error patterns. Equations (\ref{eq:12})
and (\ref{eq:21}) give us a set of equations for solving the erroneous
codeword for horizontal and vertical error patterns. When a horizontal
burst error occurs, the set of equations given by (\ref{eq:12}) can
be solved. On the other hand, the system of equations from (\ref{eq:21})
should result in no solution. If it does not happen, we have
\begin{equation}
\gamma^{i\theta}\beta^{j\phi}+\gamma^{i\theta}\beta^{(j+1)\phi}=\gamma^{k\theta}\beta^{l\phi}+\gamma^{(k+1)\theta}\beta^{l\phi},\label{eq:syn_freq_con}
\end{equation}

where, it is assumed that the syndrome for error of pattern 1 occurring
at position $(i,j)$ matches with the syndrome for error of pattern
2 occurring at position $(k,l)$. We have taken $b_{1}=2$ and $b_{2}=2.$
Thus, for disjoint syndrome set, (\ref{eq:syn_freq_con}) should have
no solution among $i,j,k,$ and $l$.

\subsubsection{Indicator Zero set}

\subsubsection*{Theorem 1 }

Let $\mathcal{Z}_{b}$ be the set of zeros of an error pattern $b(x,y)$.
Let $\mathcal{V}_{c}$ be the set of common zeros for the code and
$\mathcal{V}_{r}$ be the roots of the received codeword. If the transmitted
codeword $c(x,y)$ is affected by $b(x,y)$ and $\mathcal{Z}_{b}\cap\mathcal{V}_{c}\neq\phi$
then $\forall\left(\lambda_{x,}\lambda_{y}\right)\in\mathcal{V}_{r}$ 

\begin{equation}
\left(\lambda_{x,}\lambda_{y}\right)\in\mathcal{Z}_{b}\cap\mathcal{V}_{c}.
\end{equation}

\begin{IEEEproof}
We have, 

\begin{equation}
r(x,y)=c(x,y)+b(x,y).
\end{equation}

Let us consider $\left(\lambda_{x,}\lambda_{y}\right)\notin\mathcal{Z}_{b}\cap\mathcal{V}_{c}.$

This implies three following cases:

\textbf{Case 1:} $\left(\lambda_{x,}\lambda_{y}\right)\notin\mathcal{Z}_{b}$
and $\left(\lambda_{x,}\lambda_{y}\right)\notin\mathcal{V}_{c}.$ 

\begin{figure}[H]
\centering{}\includegraphics[draft=false, scale=0.22]{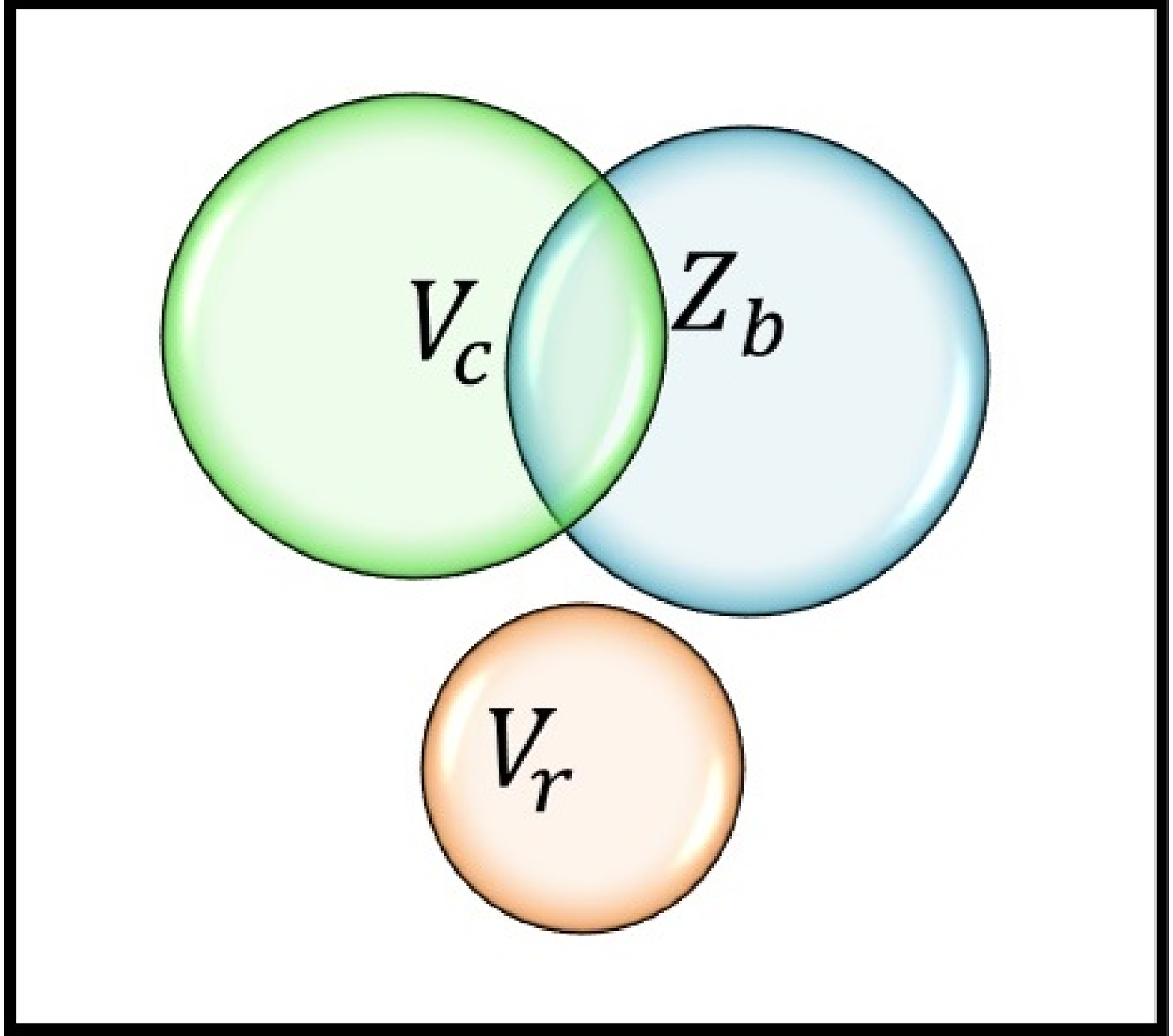}\protect\caption{$\mathcal{V}_{r}\cap\left(\mathcal{V}_{c}\cup\mathcal{Z}_{b}\right)=\phi.$}
\end{figure}

Setting $(x,y)=\left(\lambda_{x,}\lambda_{y}\right)$ we get, 

\begin{equation}
r\left(\lambda_{x,}\lambda_{y}\right)=c\left(\lambda_{x,}\lambda_{y}\right)+b\left(\lambda_{x,}\lambda_{y}\right).\label{eq:theo13}
\end{equation}

Clearly the R.H.S is not equal to zero which shows $\left(\lambda_{x,}\lambda_{y}\right)$
is not a root of $r(x,y)$, which is a clear contradiction.

\textbf{Case 2: }$\left(\lambda_{x,}\lambda_{y}\right)\in\mathcal{Z}_{b}\cap\overline{\mathcal{V}_{c}}.$ 

\begin{figure}[H]
\centering{}\includegraphics[draft=false, scale=0.22]{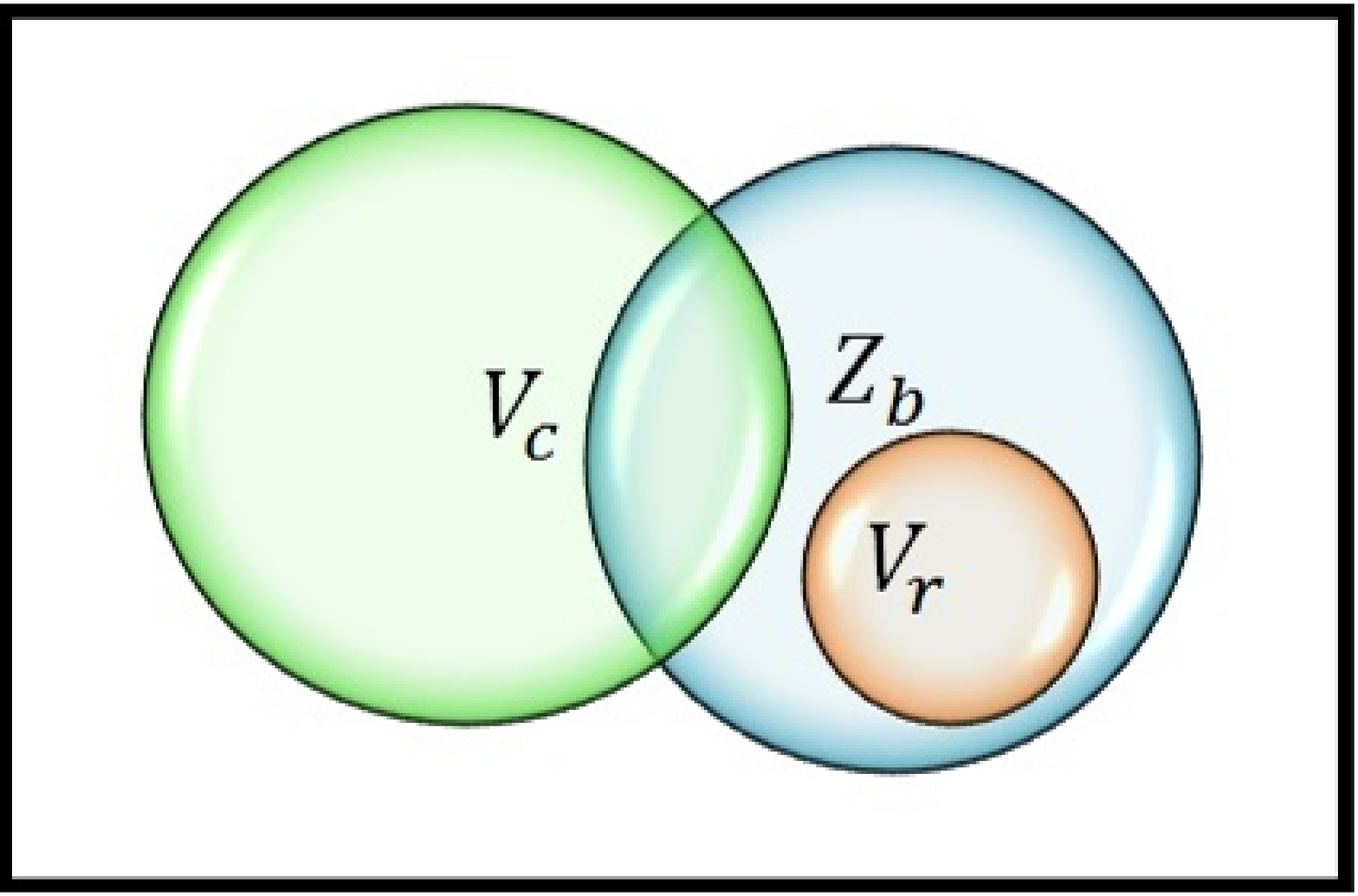}\protect\caption{$\mathcal{V}_{r}\subset\left(\mathcal{Z}_{b}\cap\overline{\mathcal{V}_{c}}\right).$}
\end{figure}

This implies $\left(\lambda_{x,}\lambda_{y}\right)\in\mathcal{Z}_{b}$
and $\left(\lambda_{x,}\lambda_{y}\right)\notin\mathcal{V}_{c}.$
Thus, we have $c\left(\lambda_{x,}\lambda_{y}\right)=0$ which implies
$\left(\lambda_{x,}\lambda_{y}\right)\in\mathcal{V}_{c}$, which is
again a contradiction.

\textbf{Case 3: }$\left(\lambda_{x,}\lambda_{y}\right)\in\overline{\mathcal{Z}_{b}}\cap\mathcal{V}_{c}.$ 

\begin{figure}[H]
\centering{}\includegraphics[draft=false, scale=0.22]{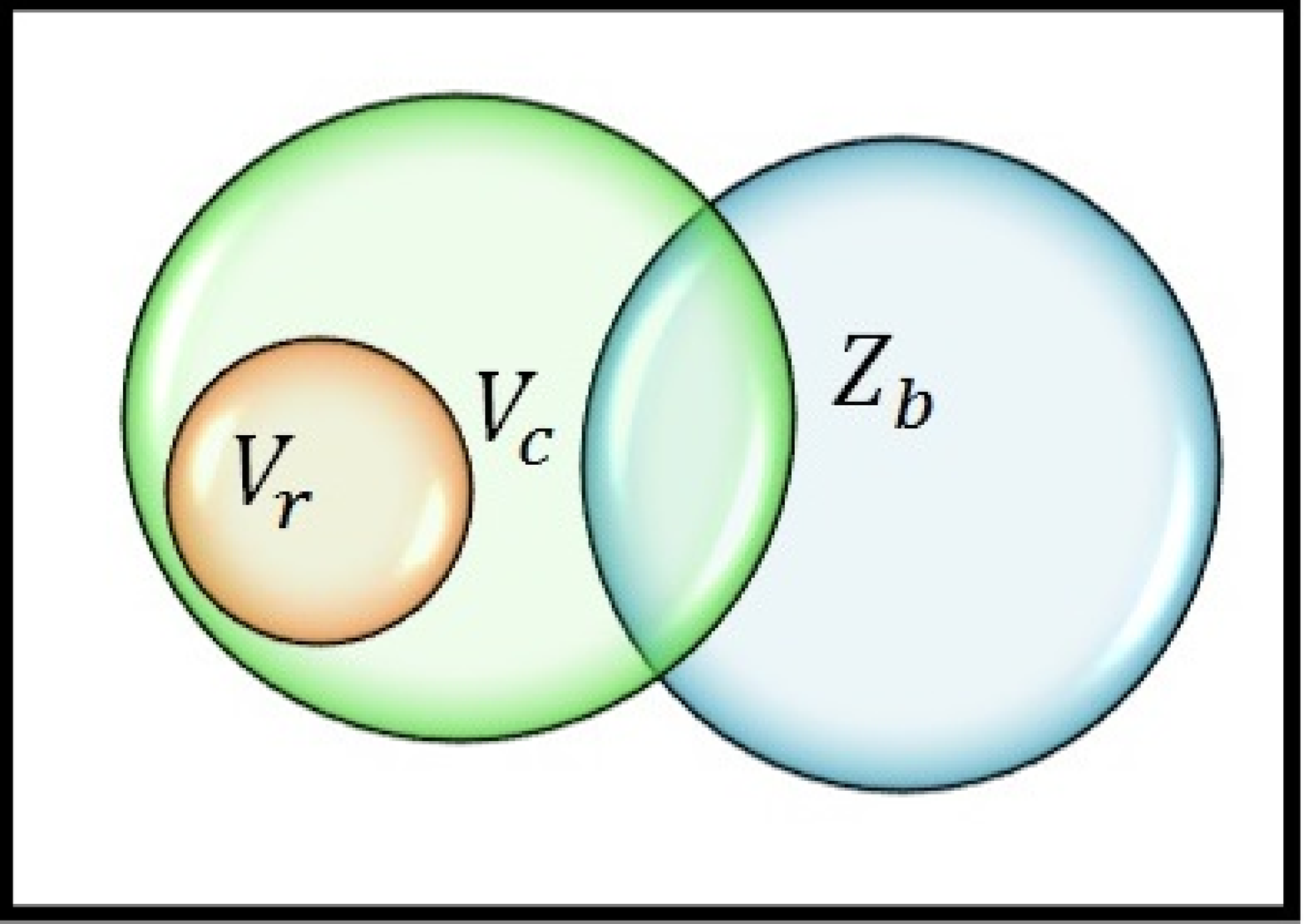}\protect\caption{$\mathcal{V}_{r}\subset\left(\overline{\mathcal{Z}_{b}}\cap\mathcal{V}_{c}\right).$}
\end{figure}

This implies $\left(\lambda_{x,}\lambda_{y}\right)\in\mathcal{V}_{c}$
and $\left(\lambda_{x,}\lambda_{y}\right)\notin\mathcal{Z}_{b}.$
Thus, we have $c\left(\lambda_{x,}\lambda_{y}\right)=0$ which implies
$\left(\lambda_{x,}\lambda_{y}\right)\in\mathcal{V}_{c}$ which is
also a contradiction.

Thus, we can conclude $\forall\left(\lambda_{x,}\lambda_{y}\right)\in\mathcal{V}_{r}$, 

\[
\left(\lambda_{x,}\lambda_{y}\right)\in\mathcal{Z}_{b}\cap\mathcal{V}_{c}.
\]

\end{IEEEproof}

\subsubsection*{Definition 3}

\textit{Indicator zero} set is a subset of the common zero set which
is used to identify the presence of an error pattern \textit{a-priori},
but not its locations. Each element of the indicator zero set is also
a root of at least one of the error pattern. 

Theorem 1 shows the necessity of including at least one zero from
each possible error pattern into the common zero set. These set of
zeros form the indicator zero set. Evaluating the received codeword
at these zero values help us to identify the error pattern. The other
zeros in the common zero set are not chosen from any set of roots
of the predefined error patterns. The syndromes at these positions
will be non-zero if the received codeword is erroneous. The non-zero
syndromes are used to construct the syndrome equation which will finally
be used to solve the error locations.

\subsubsection*{Theorem 2}

Let $\mathcal{Z}_{b}$ be the set of zeros for the error pattern $b(x,y)$,
and let $\mathcal{V}_{r}$ be the set of zeros for the received codeword
$r(x,y).$ If $\mathcal{Z}_{b}=\phi,$ then it cannot identified \textit{a-priori }and $\mathcal{V}_{r}=\phi.$
\begin{IEEEproof}
If $c(x,y)$ is the transmitted codeword affected by $b(x,y)$,
the received codeword $r(x,y)$ is given by the following equation:

\[
r(x,y)=c(x,y)+b(x,y).
\]

For all possible values $(x,y)$, we have $b(x,y)\neq0.$ Thus, $r(x,y)\neq0$, which means $\mathcal{V}_{r}=\phi$. Thus, we cannot identify such
error patterns, as there could be more than one error pattern having
empty root sets.
\end{IEEEproof}
In this paper the predefined error patterns considered are of dimension
$1\times2$ and $2\times1$.

The bi-variate polynomial expression for the 2D local error patterns
are,

\begin{eqnarray}
b_{1}(x,y) & = & 1+y,\\
b_{2}(x,y) & = & 1+x.
\end{eqnarray}

Consider a code of dimension $n\times m$. The respective set of roots
for the above equations in time domain are,

\begin{eqnarray}
\mathcal{Z}_{E}^{(1)} & = & \left\{ \left(1,1\right),\left(\gamma,1\right),\cdots,\left(\gamma^{n-1},1\right)\right\} ,\\
\mathcal{Z}_{E}^{(2)} & = & \left\{ \left(1,1\right),\left(1,\beta\right),\cdots,\left(1,\beta^{m-1}\right)\right\} .
\end{eqnarray}

Presence of both error patterns starting from different locations
and not overlapping has the following polynomial equation,

\begin{equation}
e_{3}(x,y)=x^{k_{1}}y^{l_{1}}(1+y)+x^{k_{2}}y^{l_{2}}(1+x).\label{eq:2error}
\end{equation}

The roots of this equation are given by 

\begin{equation}
\mathcal{Z}_{E}^{(3)}=\left\{ \left(1,1\right)\right\} .
\end{equation}

Thus, from Theorem 1, the common zero set should have the root $(1,1)$
to identify the presence of $e_{3}(x,y)$. To identify $e_{1}(x,y)$
and $e_{2}(x,y)$, we have to select a root from $\mathcal{Z}_{E}^{(1)}$
and $\mathcal{Z}_{E}^{(2)}$ respectively. The following set $\mathcal{V}_{c}^{(I)}$
gives us the set of indicator zeros contained in $\mathcal{V}_{c}.$

\begin{equation}
\mathcal{V}_{c}^{(I)}=\left\{ (1,1),\left(\gamma,1\right),\left(1,\beta\right)\right\} .
\end{equation}
Thus, when $r(x,y)$ is received, we evaluate it at the roots from
the set $\mathcal{V}_{c}^{(I)}$ and identify the presence of an
error pattern. We have the following cases:

\begin{flushleft}
\textbf{Case 1:} $e_{3}(x,y)$ is identified as the error pattern
if,
\par\end{flushleft}
\begin{itemize}
\item $r(1,1)=0.$
\item $r\left(\gamma,1\right)\neq0.$
\item $r\left(1,\beta\right)\neq0.$
\end{itemize}
\textbf{Case 2:} $e_{1}(x,y)$ is identified as the error pattern
if,
\begin{itemize}
\item $r(1,1)=0.$
\item $r\left(\gamma,1\right)=0.$
\item $r\left(1,\beta\right)\neq0.$
\end{itemize}
\textbf{Case 3:} $e_{2}(x,y)$ is identified as the error pattern
if,
\begin{itemize}
\item $r(1,1)=0.$
\item $r\left(\gamma,1\right)\neq0.$
\item $r\left(1,\beta\right)=0.$
\end{itemize}
\textbf{Case 4:} No error pattern is identified if,
\begin{itemize}
\item $r(1,1)\neq0,$ $r\left(\gamma,1\right)\neq0$ and $r\left(1,\beta\right)\neq0$. 
\item $r(1,1)=0,$ $r\left(\gamma,1\right)=0$ and $r\left(1,\beta\right)=0$.
\item $r(1,1)\neq0,$ $r\left(\gamma,1\right)\neq0$ and $r\left(1,\beta\right)=0.$
\item $r(1,1)\neq0,$ $r\left(\gamma,1\right)=0$ and $r\left(1,\beta\right)\neq0.$
\item $r(1,1)\neq0,$ $r\left(\gamma,1\right)=0$ and $r\left(1,\beta\right)=0.$
\end{itemize}
The last case signifies a uncorrectable error pattern. We illustrate
this through an example and formalize the procedure within Algorithm
1.

\begin{algorithm}[tbh]
\textbf{Input:} received codeword $r(x,y)$ and $\mathcal{V}_{c}$. 

\textbf{Step 1: Indicator zeros.}
\begin{itemize}
\item \textbf{Set} $\mathcal{V}_{c}^{(I)}\leftarrow\left\{ (1,1),\left(\gamma,1\right),\left(1,\beta\right)\right\} .$
\item \textbf{if} $r(1,1)=0$ \&\& $r(\gamma,1)\neq0$ \&\& $r(1,\beta)\neq0$

\begin{itemize}
\item Set $c_{1}\leftarrow1.$
\item Set $c_{2}\leftarrow1.$
\end{itemize}
\item \textbf{else if} $r(1,1)=0$ \&\& $r(\gamma,1)=0$ \&\& $r(1,\beta)\neq0$

\begin{itemize}
\item Set $c_{1}\leftarrow1.$
\item Set $c_{2}\leftarrow0.$
\end{itemize}
\item \textbf{else if} $r(1,1)=0$ \&\& $r(\gamma,1)\neq0$ \&\& $r(1,\beta)=0$

\begin{itemize}
\item Set $c_{1}\leftarrow0.$
\item Set $c_{2}\leftarrow1.$
\end{itemize}
\item \textbf{else if} $r(1,1)=0$ \&\& $r(\gamma,1)=0$ \&\& $r(1,\beta)=0$

\begin{itemize}
\item Proceed to Step 2.
\end{itemize}
\item \textbf{else }

\begin{itemize}
\item Output ``Error cannot be corrected''.
\end{itemize}
\end{itemize}
\textbf{Step 2: Solve syndrome equations}
\begin{itemize}
\item Evaluate syndrome values using the remaining values in $\mathcal{V}_{c}$on
$r(x,y).$
\item \textbf{if} all syndrome values are zero.

\begin{itemize}
\item Set $c(x,y)\leftarrow r(x,y).$
\item Go to \textbf{Output}.
\end{itemize}
\item \textbf{else }

\begin{itemize}
\item Solve the syndrome equations.
\item Flip the solved erroneous positions to get $c(x,y).$
\item Go to \textbf{Output.}
\end{itemize}
\end{itemize}
\textbf{Output:} $c(x,y)$.

\protect\caption{Decoding Algorithm }
\end{algorithm}

\subsubsection*{Example 2}

Let us assume that we have transmitted the zero codeword. Our basic
error patterns are $1\times2$ and $2\times1$. Let the common zero
set in the frequency domain be, 

\begin{equation}
\mathcal{V}{}_{c}=\{(1,1)(\gamma,1),(1,\beta),(\gamma,\beta),(\gamma^{2},\beta^{3})\}.
\end{equation}
 Consider the following two received codewords.

\[
\begin{array}{cc}
r_{1}=\left(\begin{array}{ccccc}
1 & 1 & 1 & 0 & 0\\
0 & 0 & 1 & 0 & 0\\
0 & 0 & 0 & 0 & 0
\end{array}\right) & r_{2}=\left(\begin{array}{ccccc}
0 & 0 & 0 & 0 & 0\\
0 & 0 & 1 & 1 & 0\\
0 & 0 & 0 & 0 & 0
\end{array}\right).\end{array}\text{ }
\]

The syndrome equation considered here is given by the following equation,

\begin{eqnarray}
E_{\theta,\phi} & = & c_{1}\left(\gamma^{k_{1}\theta}\beta^{l_{1}\phi}+\gamma^{k_{1}\theta}\beta^{(l_{1}+1)\phi}\right)+\nonumber \\
 &  & c_{2}\left(\gamma^{k_{2}\theta}\beta^{l_{2}\phi}+\gamma^{(k_{2}+1)\theta}\beta^{l_{2}\phi}\right).
\end{eqnarray}

The elements $(0,0)$, $(0,1)$ and $(1,0)$ are the indicator zeros.

\subsubsection*{Decoding $r_{1}$}

We have,

\begin{equation}
r_{1}(x,y)=1+y+y^{2}+xy^{2}.
\end{equation}

It is clear that $r_{1}(1,1)=0,$ $r_{1}(\gamma,1)=\alpha^{10}$ and
$r_{1}(1,\beta)=\alpha^{14}.$ Thus, the error polynomial is of the
form equivalent to $e_{3}(x,y).$ Hence, we make $c_{1}=1$ and $c_{2}=1.$
To solve for the error locations, we plug in $(\gamma,1),\text{ }(1,\beta),\text{ }(\gamma,\beta),\text{ and }(\gamma^{2},\beta^{3})$
in equation (\ref{eq:2error}) to obtain the following syndrome equations:

\begin{eqnarray*}
\gamma^{k_{2}}+\gamma^{(k_{2}+1)} & = & \alpha^{10},\\
\beta^{l_{1}}+\beta^{(l_{1}+1)} & = & \alpha^{14},\\
\gamma^{k_{1}}\beta^{l_{1}}+\gamma^{k_{1}}\beta^{(l_{1}+1)}+\gamma^{k_{2}}\beta^{l_{2}}+\gamma^{(k_{2}+1)}\beta^{l_{2}} & = & \alpha^{7},\\
\gamma^{2k_{1}}\beta^{3l_{1}}+\gamma^{2k_{1}}\beta^{3(l_{1}+1)}+\gamma^{2k_{2}}\beta^{3l_{2}}+\gamma^{2(k_{2}+1)}\beta^{3l_{2}} & = & \alpha^{3}.
\end{eqnarray*}

Solving the above equations, we have $\left(k_{1,}l_{1}\right)=(0,0)$
and $\left(k_{2,}l_{2}\right)=(0,2)$.

\subsubsection*{Decoding $r_{2}$}

We have,

\begin{equation}
r_{1}(x,y)=xy^{2}+xy^{3}.
\end{equation}

It is clear that $r_{1}(1,1)=0,$ $r_{1}(\gamma,1)=0$ and $r_{1}(1,\beta)=\alpha^{5}.$
Thus, the error polynomial is of the form equivalent to $e_{1}(x,y).$
Hence, we make $c_{1}=1$ and $c_{2}=0.$ To solve for the error locations,
we plug in $(1,\beta),\text{ }(\gamma,\beta),\text{ and }(\gamma^{2},\beta^{3})$
in $e_{1}(x,y)=x^{k_{1}}y^{l_{1}}(1+y)$ to obtain the following syndrome
equations:

\begin{eqnarray*}
\beta^{l_{1}}+\beta^{(l_{1}+1)} & = & \alpha^{5},\\
\gamma^{k_{1}}\beta^{l_{1}}+\gamma^{k_{1}}\beta^{(l_{1}+1)} & = & \alpha^{10},\\
\gamma^{2k_{1}}\beta^{3l_{1}}+\gamma^{2k_{1}\theta}\beta^{3(l_{1}+1)} & = & \alpha^{5}.
\end{eqnarray*}

Solving the above equation, we have $\left(k_{1,}l_{1}\right)=(1,2).$

\subsubsection*{Proposition1}
Presence of overlapping predefined error patterns of dimension $1\times2$ and $2\times1$ cannot be determined \textit{a-priori} and hence, cannot be corrected. 
\begin{IEEEproof}
The polynomial expression of the error pattern as a linear combination of the error pattern of dimension $1\times2$ and $1\times2$ at position $\left(k_{1,}l_{1}\right)$ and $\left(k_{2,}l_{2}\right)$ respectively is given by,
\begin{equation}
e(x,y)=x^{k_{1}}y^{l_{1}}(1+y)+x^{k_{2}}y^{l_{2}}(1+x).
\end{equation}

Without loss of generality, let $k_{2}=k_{1}$ and $l_{2}=l_{1}+1$ for overlapping errors. We get,
\begin{equation}
e(x,y)=x^{k_{1}}y^{l_{1}}+x^{k_{1}+1}y^{l_{1}+1},
\end{equation}
which gives two isolated errors. Such an overlapping burst can be treated as a disjoint combination of an error of dimension $1\times2$ error and an isolated error, or as a disjoint combination of an error of dimension $2\times1$ and an isolated error. From Theorem 2, we have established that an isolated error is not within our set of predefined error patterns. Thus, an error pattern formed in combination of any valid predefined error pattern with an isolated error cannot be identified \textit{a-priori}.
\end{IEEEproof}

\subsection{Multiple Bursts of Same Type}

If the total number of bits in error is within the minimum distance
bound, this code can be used to correct the presence of multiple bursts
of an error of a particular type. Let us assume that the bursts are non-overlapping.
For the general case, suppose there are $\mu$ number of burst errors
of type $1\times\lambda$ that are non-overlapping, we can uniquely
identify the position of the errors. The following theorem summarizes this result.

\subsubsection*{Theorem 3}

Presence of $\mu$ number of burst errors of type $1\times\lambda$
can be uniquely identified. 
\begin{IEEEproof}
Presence of $\mu$ number of burst errors of type $1\times\lambda$
leads to the following expression in the transformed domain:
\begin{equation}
\sum_{\sigma=0}^{\mu-1}\left(\sum_{\omega=0}^{\lambda-1}\gamma^{i_{\sigma}\theta}\beta^{(j_{\sigma}+\omega)\phi}\right)=E_{\theta,\phi}.\label{eq:mul1}
\end{equation}
Equation (\ref{eq:mul1}) can be simplified further as
\begin{equation}
\sum_{\sigma=0}^{\mu-1}\gamma^{i_{\sigma}\theta}\beta^{j_{\sigma}\phi}=\frac{E_{\theta,\phi}}{\sum_{\omega=0}^{\lambda-1}\beta^{\omega\phi}}.
\end{equation}

Let $\{(\zeta{}_{\sigma},\eta_{\sigma})\}$ and $\{(\zeta'{}_{\sigma},\eta'_{\sigma})\}$
for $0\leq\sigma\leq\mu-1$ be two solutions of the above equation
with none of component of the solutions equal. As the right hand side
of the above equation does not have a variable, we have 
\begin{equation}
\sum_{\sigma=0}^{\mu-1}\gamma^{\zeta{}_{\sigma}\theta}\beta^{\eta{}_{\sigma}\phi}+\sum_{\sigma=0}^{\mu-1}\gamma^{\zeta'_{\sigma}\theta}\beta^{\eta'_{\sigma}\phi}=0.\label{eq:unique_th2}
\end{equation}

Equation (\ref{eq:unique_th2}) implies either the terms are all equal,
or they are pairwise equal. This gives us the following ordered pairs:
\begin{eqnarray}
(\zeta{}_{0},\eta_{0}) & = & (\zeta'{}_{0},\eta'_{0}),\nonumber \\
(\zeta{}_{1},\eta_{1}) & = & (\zeta'_{1},\eta'_{1}),\nonumber \\
\cdots & \cdots & \cdots\nonumber \\
(\zeta{}_{\mu},\eta_{\mu}) & = & (\zeta'_{\mu},\eta'_{\mu}).
\end{eqnarray}

Any other combination will just be a permutation of the solution set
which means, the solution set is unique, proving the result. 
\end{IEEEproof}
The following example will illustrate our idea.

\subsubsection*{Example 3}

Let the received codeword be 
\[
r=\left(\begin{array}{ccccc}
0 & 0 & 0 & 1 & 1\\
0 & 0 & 0 & 0 & 0\\
1 & 1 & 0 & 0 & 0
\end{array}\right).
\]

Let the two bursts start at positions $(i,j)$ and $(k,l)$. The corresponding
error polynomial is
\begin{equation}
E_{\theta,\phi}=\gamma^{i\theta}\beta^{j\phi}+\gamma^{i\theta}\beta^{(j+1)\phi}+\gamma^{k\theta}\beta^{l\phi}+\gamma^{k\theta}\beta^{(l+1)\phi}.
\end{equation}

Let the common zero in the frequency domain be
\begin{equation}
\mathcal{VF}_{c}=\{(0,0),(1,1),(1,4),(2,2),(2,3)\}.
\end{equation}
This gives us the following equations as before i.e.,
\begin{eqnarray*}
E_{0,0} & = & 0=R_{0,0},\\
E_{1,1} & = & \gamma^{i}\beta^{j}+\gamma^{i}\beta^{(j+1)}+\gamma^{k}\beta^{l}+\gamma^{k}\beta^{(l+1)}=\alpha^{12},\\
E_{1,4} & = & \gamma^{i}\beta^{4j}+\gamma^{i}\beta^{4(j+1)}+\gamma^{k}\beta^{4l}+\gamma^{k}\beta^{4(l+1)}=\alpha^{3},\\
E_{2,2} & = & \gamma^{2i}\beta^{2j}+\gamma^{2i}\beta^{2(j+1)}+\gamma^{2k}\beta^{2l}+\gamma^{2k}\beta^{2(l+1)}=\alpha^{9},\\
E_{2,3} & = & \gamma^{2i}\beta^{3j}+\gamma^{2i}\beta^{3(j+1)}+\gamma^{2k}\beta^{3l}+\gamma^{2k}\beta^{3(l+1)}=\alpha^{6}.
\end{eqnarray*}

The solution to the above system of equations gives the result $(i,j)=(0,3)$
and $(k,l)=(2,0)$ which are exactly the positions where the $1\times2$
bursts start.

Previous decoding schemes \cite{key-8} required storing the syndrome polynomials
for each error pattern. In our case, the erroneous positions are found
directly by solving the appropriate syndrome equations.

\section{Conclusions }

We presented a 2D cyclic code construction in the frequency domain.
Our code is capable of correcting simultaneous occurrence of multiple predefined error
patterns. The encoding and decoding steps are done entirely in the transformed domain, thereby, facilitating efficient realization. Further work is needed to extend this to a broad framework of 2D codes over curves.

\section*{Acknowledgment}

The authors would like to thank the Department of Electronics and
Information Technology, grant no. MITO0101, for supporting this work.

\end{document}